\documentclass[conference]{IEEEtran}
\IEEEoverridecommandlockouts
\usepackage{cite}
\usepackage{amsmath,amssymb,amsfonts}
\usepackage{algorithmic}
\usepackage{graphicx}
\usepackage{textcomp}
\usepackage{xcolor}
\usepackage{url}
\usepackage{hyperref}
\usepackage{soul}
\def\BibTeX{{\rm B\kern-.05em{\sc i\kern-.025em b}\kern-.08em
    T\kern-.1667em\lower.7ex\hbox{E}\kern-.125emX}}
\begin{document}

\title{A Scalable Framework for Post-Quantum Authentication in Public Key Infrastructures
\thanks{This work was partially supported by the EU project ALLEGRO, GA-ID101092766.}
}

\author{
\IEEEauthorblockN{ Antonia Tsili }
\IEEEauthorblockA{
\textit{National and Kapodistrian University of Athens, Greece} \\
tonyts@di.uoa.gr \\
https://orcid.org/0000-0001-6959-5338
}
\and
\IEEEauthorblockN{ Konstantinos Kordolaimis \hspace{0.1\linewidth}}
\IEEEauthorblockA{\textit{Eulambia Advanced Technologies, Greece} \hspace{0.1\linewidth}\\
sdi2000091@di.uoa.gr \hspace{0.1\linewidth}}
\and  
\IEEEauthorblockN{ \hspace{0.105\linewidth} Konstantinos Krilakis }
\IEEEauthorblockA{\hspace{0.105\linewidth} \textit{Eulambia Advanced Technologies, Greece}\\
 \hspace{0.105\linewidth} konstantinos.krilakis@eulambia.com 
 }
\and
\IEEEauthorblockN{Dimitris Syvridis}
\textit{National and Kapodistrian University of Athens, Greece}\\
\textit{Eulambia Advanced Technologies, Greece} \\
dsyvridi@di.uoa.gr \\
https://orcid.org/0000-0001-8026-576X
}

\maketitle

\begin{abstract}
This work explores the performance and scalability of a hierarchical certificate authority framework with automated certificate issuance employing post-quantum cryptographic (PQC) signature algorithms. The system is designed for compatibility with both classical and PQC algorithms, promoting crypto-agility while ensuring robust security against quantum-based threats. The proposed framework design expects minimal cryptographic requirements from potential clients, protects certificates of high importance against cross-dependent chains-of-trust and allows for prompt switching between classical and PQC algorithms. Finally, we evaluate SPHINCS$^+$, Falcon, and Dilithium variants in various configurations of certificate issuance and verification accommodating a large client base, underlining the trade-offs in balancing performance, scalability, and security.
\end{abstract}


\section{Introduction}
While the concept of quantum computers was introduced less than a lifetime away from the invention of the first electronic computer, their potential has recently alarmed the cryptographic community. The proven robustness of classical cryptography algorithms, among which RSA~\cite{RSA} holds a prominent position, has resulted into their widespread adoption. Such algorithms have received limited improvements, but their backbone adheres to similar mathematical guarantees. The development of Shor's algorithm~\cite{shor1994} has, therefore, played a decisive role in the recent advance of cryptography, showing that classical cryptographic methods are vulnerable to attacks from quantum computers. 

The concern is underlined by the National Institute of Standards and Technology (NIST), a U.S. federal agency playing a pivotal role in cryptography by establishing cryptographic standards and guidelines to ensure secure communication and data protection for government and industry. In view of security attacks from quantum computers, NIST in 2016 initiated the concentration, trialing, assessment and standardization of post-quantum cryptography (PQC) algorithms. These algorithms are developed within the scope of mathematical foundations that avoid the vulnerability of classical algorithms, in view of their widespread adoption. In addition to PQC, the most promising mitigation practices also include quantum cryptography, especially Quantum Key Distribution (QKD)~\cite{QKDnets, Wang2021, Cao2022}, which can be efficiently complemented by the PQC-focused work presented here.

PQC methods are meant to replace the classical cryptography methods; however, this transition has become a burning subject of research, due to the need for adoption of new requirements connected to their properties. In particular, PQC algorithms present increased needs of memory and computation power, giving rise to the problem of cryptographic agility (crypto-agility)~\cite{ mattsson2023proposals, PaulNiethammer, ELCA, marchesi2024}, defined as their widespread implementation to existing networks. Research is mainly focused on the incorporation of four NIST standardization finalists, including CRYSTALS-Kyber for key establishment~\cite{Kyber}, CRYSTALS-Dilithium~\cite{Dilithium}, Falcon~\cite{prest2020falcon} and SPHINCS$^+$~\cite{Sphincs} for digital signatures. In fact, NIST announced three finalized post-quantum safe Federal Information Processing Standards~\cite{NIST-FIPS-203,NIST-FIPS-204,NIST-FIPS-205} on 13th August 2024 based on CRYSTALS-Kyber, CRYSTALS-Dilithium and SPHINCS$^+$, placing the latter two under the spotlight.

The current work explores the implementation of PQC in the context of authentication, especially a public key infrastructure (PKI). We create a framework that can assist in readily transitioning current PKI topologies to the post-quantum era. PKI refers to a system that functions through the management and usage of certificates. The certificates are signed by well-known and trusted Certificate Authorities (CA), whose digital signature is produced using its private key and can be verified using its public key. An entity that owns a certificate bearing such a verified signature is considered secure, since it traces back to the trusted CA that lent its signature. This way, different entities and devices over the network can authenticate themselves, proving their identity through the signed certificate. Our framework consists of the usual chain-of-trust scheme, including a root CA and intermediate CAs (ICAs). Clients, also referred to as end-entities (EEs), may request the issuance of their certificates from ICAs, which, in turn, obtain their certificates from the root Certificate Authority (CA). 
We propose a holistic approach to the certificate authority model, that considers the functionality of different components towards security, data flow, maintenance, independence and accessibility. This prototype sketches and safeguards the higher levels of the certificate hierarchy against compromise of some chain-of-trust, while keeping processing quick. For example, keeping multiple roots, as recommended here, is a crucial feature. After all, management of the co-existence of classical and PQ cryptography needs radical design alterations at the core, the root CA, in order to preserve security and introduce cryptographic agility.

This work makes the following key contributions:
\begin{itemize}
    \item \textbf{Scalable certificate authority employing PQC signatures.} We create a certificate authority which securely manages different chains-of-trust for a large client base, offering certificates signed with PQC signature algorithms.
    \item \textbf{Classical and PQC compatibility and adaptability.} Our framework is designed in view of accommodating both new and classical algorithms on-the-fly at the critical levels of certificate authorities.
    \item \textbf{Automated certificate issuance with PQC signatures.} We introduce a transferable framework that enables the automated issuance and verification of digital certificates using PQC signature algorithms, ensuring robust security against quantum-based threats.
    \item \textbf{Hierarchical three-Layer architecture on a full-fledged network simulation.} The proposed system is structured into three distinct layers -- CA, ICA, and EEs --facilitating efficient certificate management and scalability. The scheme targets the application to QKD domains, where QKD nodes represent ICAs that can easily mutually authenticate and distribute certificates to their clients.
    \item \textbf{Performance metrics for operations including certificates.} We provide detailed evaluations of the framework's performance, including metrics for signing, verifying, and distributing certificates across multiple clients, with support for up to two ICAs in the hierarchy.
    \item  \textbf{Crypto-agility promotion.} Our design enhances crypto-agility by allowing the use of PQC signature algorithms at the CA and ICA levels, without any PQC requirements at the client level, ensuring backward compatibility and ease of adoption even in IoT.
\end{itemize}

\section{Existing related Work}

The relatively new field of PQC is currently being explored in many aspects, ranging from theoretical security proofs, to practical guidelines and preliminary application testbeds. Apropos of recent work on PQC incorporation to PKI, Bene and Kiss in~\cite{Fruzsina2023} have provided an overview of basic PKI concepts, as well as a description of the main PQC algorithms and methods that can be integrated with PKI. Further on this topic, Paul S. et al.~\cite{MixedChains2022} have proposed a 2-step PQC migration strategy and have demonstrated a preliminary application of it to three commonly implemented layers -- the root CA, the ICAs and the EEs. Their two-step migration approach enables a smooth transition to PQC authentication, wherein the first step comprises mixed certificate chains, rooted in PQC signatures but incorporating classical cryptography at higher PKI levels. This strategy aims to protect critical information, like root CA keys and certificates, during the ongoing transition. We adopt these recommendations to develop a cloud-based framework that enhances crypto-agility and accelerates large-scale implementation.

Contrary to Kampanakis and Kallitsis in~\cite{Kampanakis2022}, our current work motivates the inclusion of ICAs in the network topology. This choice promotes easier multiple EEs' management, security --by decoupling the root CA from EEs in a chain-of-trust-- and faster processing, since certificate distribution is carried out by a group of CAs.

The size of PQ keys and certificates has sparked numerous discussions regarding their compatibility with existent and widespread protocols, raising concerns about increased overheads, memory requirements and reduced security. 

Kampanakis and Kallitsis in~\cite{Kampanakis2022}, propose the suppression of ICA involvements in PQ TLS handshakes in order to avoid the overhead added by verifying the whole certificate chain. Their solution follows the ``supply and demand'' aspect of certificate dependence, which means that they address the issue of the certificate chain by storing the ICA certificates that have issued the certificates of the peers of interest. This static approach reduces the data volume to communicate between peers and, consequently, reduces the overhead of message exchange, however, needs further investigation in terms of application requirements and guarantees, as well as more large-scale adaptations.

The concept of automated certificate issuance in classical network configurations has been primarily a concern of well-known certificate authorities, whose operation is instrumented by the Automatic Certificate Management Environment (ACME) protocol~\cite{rfc8555}. Shifting to the post-quantum era, Giron et al.~\cite{Giron2024} have studied the integration of PQC into PKI through the adaptation of the ACME. Their work is focused on the automated issuance of client PQC certificates from the ACME server bearing the Let’s Encrypt Client and the proposal of a new challenge as a part of the issuance process. While this approach provides some intuition on the impact of PQC to the standardized certification automation, we create the certificate authority from scratch and examine the automated PQC certificate issuance process of a large scale hierarchical PKI.

\section{Post-Quantum authentication services} \label{sec:system}
We hereafter present our scalable and extensible framework along with some experimental reproducible results related to its function in the employment of PQC signature algorithms. 
The proposed framework comprises a certificate authority and an automatic certificate issuance system able to readily incorporate PQC algorithms to existing networks. The certificates produced reflect the implemented hierarchy, which is widely adopted, 
yet the different entities' roles are remodeled. 
We make a start in post-quantum readiness for authentication, including that of QKD nodes', by preparing a system with services that can manage and monitor the respective certificates. 
Ultimately, the existing QKD systems have their own simplified key management system (ETSI014\cite{etsi_qkd_014} in most existing devices) and a novice authentication system. In fact, certification is a major missing fundamental functionality of the existing QKD networks, therefore, our system can easily fill this gap with proper adaptations.
The realization of the authentication among peers can then be addressed through the QKD server or defining domains on physical layer, left for future work.
We will initially present the roles of the entities designed to participate in the proposed framework, followed by the description of their implementation in our experimental setup and the analysis of our experiments in Section~\ref{sec:setup} and Section~\ref{sec:results} respectively.

\begin{table*}[htbp]
\caption{List of PQC algorithms used with the corresponding NIST security level, public key size, private key size and signature size.}
\begin{center}
\begin{tabular}{|c|c|c|c|c|}
\hline
\multicolumn{5}{|c|}{\textbf{PQC Signature Algorithms}} \\
\cline{1-5} 
Algorithm & \multicolumn{1}{l|}{Claimed NIST Level} & \multicolumn{1}{l|}{Public key (bytes)} & \multicolumn{1}{l|}{Secret key (bytes)} & \multicolumn{1}{l|}{Signature (bytes)} \\ \hline
falcon512  & 1 & 897  & 1281 & 752  \\ \hline
falcon1024 & 5 & 1793 & 2305 & 1462 \\ \hline
Dilithium2  & 2 & 1312 & 2528 & 2420 \\ \hline
Dilithium3 & 3 & 1952 & 4000 & 3293 \\ \hline
Dilithium5 & 5 & 2592 & 4864 & 4595 \\ \hline
SPHINCS+-SHA2-128f-simple  & 1 & 32 & 64 & 17088 \\ \hline
SPHINCS+-SHA2-192f-simple  & 3 & 48 & 96 & 35664 \\ \hline
\end{tabular}
\end{center}
\label{tab:PQCsizes}
\end{table*}

\subsection{Root CA}
\begin{figure}[htbp]
    \centerline{\includegraphics[width=.5\textwidth]{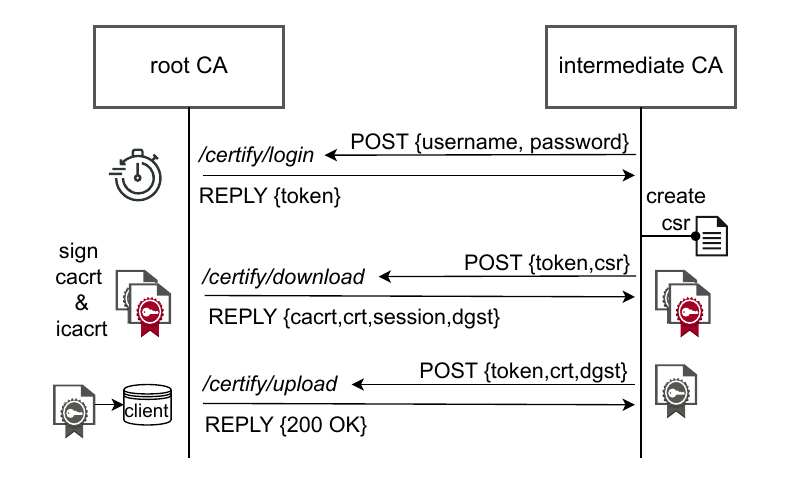}}
    \caption{Interactions between the root CA and an intermediate CA. The root CA endpoints are reached in the context of the \textit{certify} service using HTTP requests.}
    \label{fig:CAends}
    \end{figure}

The root CA activity is distributed into three services aimed at serving the intermediate CAs' requests. EEs are not included in the list of intermediate CAs held by the root CA, hence service is declined by default. Overall, the clients of these services are allowed to be registered entities which are destined to own intermediate CAs' certificates. EEs are never given access to CA's certificate and public key. The services are distinguished into enrollment, certification and verification and are exposed through different endpoints that correspond to different steps of the certification process, as shown in \textbf{Fig.~\ref{fig:CAends}}. 

    \paragraph{enroll} 
    The enrollment service exposes an \textit{/enroll} endpoint where the intermediate CA \textsc{POST}s the entity's information that has already been  communicated to the root CA. The service subsequently generates a new certificate specifically generated for the current client, ensuring that the chain of certificates derived from the root CA remains distinct and independent from other client's chains. This approach preserves the integrity and independence of the original certificate path linked to the root CA. 
    \paragraph{certify} 
    Before accessing the main certificate issuance service endpoints, the client produces a certificate signing request (CSR) using their own public key. The main certificate issuance service includes three steps; the first step is served through the \textit{/certify/login} endpoint, where the client \textsc{POST}s their own information to get a timed token. The token helps in raising defense against replay attacks~\cite{stamp2011information} and is necessary for requesting service through the main endpoints. Afterwards, the client \textsc{POST}s a request that contains the client's token and CSR to \textit{/certify/download}. If the token is valid, the root CA responds with the root CA and intermediate CA signed certificates, which are added to the intermediate CA's database. The intermediate CA's certificate is sent back to the root CA as a message of acknowledgement and for monitoring purposes through a \textsc{POST} request to the \textit{/certify/upload} endpoint.
    
    \begin{figure}[htbp]
    \centerline{\includegraphics[width=.5\textwidth]{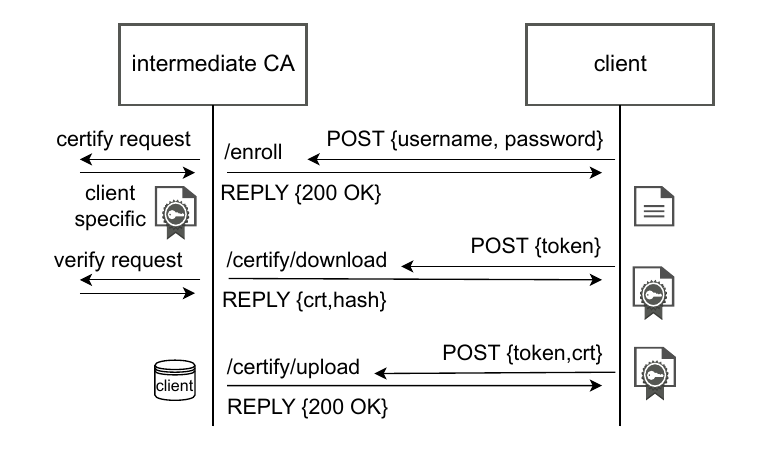}}
    \caption{Interactions between an intermediate CA (ICA) and the EE (client). The ICA endpoints are reached in the context of the  \textit{enroll} and \textit{certify} service, while the root CA endpoints are reached in the context of the \textit{verify} service, using HTTP requests.}
    \label{fig:ICAends}
    \end{figure}
    \paragraph{verify} 
    The verification service is available through the \textit{/verify/check} endpoint, once the \textit{/verify/login} endpoint has granted access by providing a token to the enrolled client. This is used to validate the certificate chain, along with the embedded signatures, when an EE requests the issuance of a certificate. It is mainly used as a confirmation message from the root CA that the intermediate CA is, in various aspects, entitled to sign certificates. 

\subsection{Intermediate CA}
The intermediate CA server activity concerns EE clients and offers services similar to the root CA. In other words, it includes the \textit{/enroll}, \textit{/certify/login}, \textit{/certify/download}, \textit{/certify/upload}, \textit{/verify/login} and \textit{/verify/check} endpoints, however the clients have different permissions compared to the interaction of root CA and ICA. The interactions of clients and ICAs are shown in \textbf{Fig.~\ref{fig:ICAends}}.

    \paragraph{enroll} 
    The enrollment service exposes an \textit{/enroll} endpoint where the EE \textsc{POST}s the entity's information. The intermediate CA subsequently follows the certification process by creating a certificate signing request (CSR) and applying for a new certificate to the root CA, which will be used for creating the EE's certificate chain.
    \paragraph{certify} 
    Before accessing the main certificate issuance service endpoints, the client produces a CSR using their own public key. The main certificate issuance service includes the same three steps as before; the first step is served through the \textit{/certify/login} endpoint, where the client \textsc{POST}s their own information to get a timed token. Afterwards, the client \textsc{POST}s a request that contains the client's token and CSR to \textit{/certify/download}. If the token is valid, the intermediate CA validates its own certificate chain by accessing the verify service of the root CA. If verified, the intermediate CA responds with the client's signed certificate. The EE's certificate is sent back to the intermediate CA through a \textsc{POST} request to the \textit{/certify/upload} endpoint.
    \paragraph{verify} 
    A verification service is also available for EEs through the \textit{/verify/check} endpoint, once the \textit{/verify/login} endpoint has granted access by providing a token to the enrolled client. It can be used from the EE each time its certificate is accessed in the context of another application. 

\subsection{EE clients}
The EEs referred to as clients interact with the ICA services and are deployed on multiple devices. Specifically, they are implemented as workflows that POST requests to ICA's \textit{/enroll} along with their credentials and, subsequently, request certificate issuance. This is done by POSTing requests to the ICA's \textit{/certify/download} and \textit{/certify/upload}, after they have logged into the service at \textit{/certify/login}, much like the interactions between ICA and CA. 
\bigskip

The system effectively mimics the diverse and concurrent interactions of a large-scale client base, providing a realistic environment to test and validate the functionality and scalability of the proposed framework. Each ICA is uniquely responsible for its clients, and the root CA manages the ICAs, distributing management between the different PKI layers and promoting security and scalability. Details are described in Section~\ref{sec:details}.

\section{Description of experiments} \label{sec:edescr}

Our framework is formed using containerized applications which involve cryptography providers. The providers constitute modules that offer a set of cryptographic libraries, in this case including classical and post quantum algorithms. With just a couple runtime  adjustments within the containerized applications of the different CA levels, the circulated certificates can be signed with either classical or PQC algorithms on the corresponding level, a feature that will later be automated within security limitations.
We test our system by performing two sets of experiments on the employment of PQC signature algorithms for certificate issuance. The results from the two sets of experiments highlight the impact of PQC adoption at different levels of the PKI, but also empirically reflect the 2-step migration to PQC strategy proposed in~\cite{MixedChains2022}. 

In the first set of experiments, different PQC signature algorithms are employed at the CA level only. We perform our tests using the Dilithium, SPHINCS$^+$ and Falcon variants shown in \textbf{Table~\ref{tab:PQCsizes}}, the first two being the foundations for the formulation of the recent NIST-approved standards FIPS 204~\cite{NIST-FIPS-204} and 205~\cite{NIST-FIPS-205}, while Falcon is an efficient alternative currently being processed to constitute the basis of another NIST-approved standard. The different variants produce different key sizes, and, as a result, different certificate sizes, and correspond to different security levels~\cite{nist80057, lenstra2004}. In these experiments, the system comprises the root CA and two different ICAs located at Machine 2 and Machine 3 of Table~\ref{tab:hardware} respectively. Each ICA receives requests from 25,50,100,250 and 500 natively located clients concurrently, and then produces the matching requests addressed to the root CA according to the workflow described in Section~\ref{sec:system}.

In the second set of experiments the same set of PQC signature algorithms is used. Each algorithm is employed both at the ICA level, as well as the CA level. In these experiments, the ICA is placed at Machine 1 and clients at Machine 2 and Machine 3 of Table~\ref{tab:hardware} . The ICA receives requests from 50,100,200,500 and 1000 remotely located clients concurrently, and then produces the matching requests addressed to the root CA.

In both sets of experiments, the PKI levels where PQC algorithms are absent use the vastly adopted RSA with SHA256 for key generation and/or signature creation, which produces a 2048-bit public/secret key. This is chosen so that the framework is tested against a classical algorithm with a large load, contrary to the case in which ECDSA is used. The certificates produced follow the X.509 format and are encoded in PEM, while the information load is consistent among experiments and minimal. This helps focusing on metrics associated with network loads and process times independently to the CSR/certificate size. The results of the experiments are discussed in the next.

\section{Experimental setup} \label{sec:setup}
The designed software components constitute our framework's backbone, being supported by the hardware components listed in \textbf{Table~\ref{tab:hardware}}. Our software components are designed to communicate over the Internet using simple HTTP requests and are easily transferable. In this work, the topology implemented for conducting our experiments is designed to follow a Full-Cloud-Fog architecture~\cite{javadpour2023encryption}, described in Section~\ref{sec:topology}. Implementation details regarding the software and hardware components of the setup are elaborated on in Section~\ref{sec:details}.

\subsection{System Architecture} \label{sec:topology}

The proposed topology in this work is organized over a three-layer scheme encompassing cloud, fog and client nodes. Within this structure, cryptographic and management related tasks are distributed among the cloud and fog nodes, effectively balancing computational demands between these two layers. This approach reduces computational complexity on edge devices, however, it is aligned with the principles of edge computing by situating certain processing tasks closer to the data sources. In the presented topology, the top layer consists of a remote server that acts as the cloud node and monitors the activities of the fog nodes. The middle layer comprises fog nodes that serve as domain-local gateway devices. The fog nodes are responsible for managing the device nodes within their respective domains, which act as clients to the fog nodes. The network is described by a hybrid star and hierarchical topology:
\begin{enumerate}
    \item \textit{Server-Router Link:} The remote server $\mathit{S}$ is accessed through a central router $\mathit{R}$.

    \item \textit{Router-Fog Nodes Links:} The central router $R$ is connected to multiple fog (QKD) nodes $\mathit{Q_1,Q_2,...,Q_n}$ representing the corresponding domains $\mathit{D_1,D_2,...,D_n}$ . In the case of QKD nodes, the classical channel is complemented by the quantum channel connecting the QKD nodes directly for key agreement through quantum mechanics operations.

    \item  \textit{QKD Nodes-End Devices Links:} Each fog node $\mathit{Q_i}$ is connected to a set of end devices $\mathit{E_{i1},E_{i2},...,E_{im}}$. The end devices rely on the fog nodes for cryptographic services such as authentication or encryption.
\end{enumerate}
Each layer mentioned above serves a distinct function in the automated certification issuance process, with different tasks and permissions defined at the top level. The certificate hierarchy and process flow are closely linked to the software components developed that were distributed among the hardware components, all of which will be described in Section~\ref{sec:details}.

\subsection{Implementation details} \label{sec:details}

Our implementation involves various software components that constitute the backbone of a scalable automated certification issuance system. We included different hardware components that we connected over a real-world physical network and distributed the software components accordingly. The setup was tested on three separate devices. The implementation involves three main software entities that were distributed among the physical devices connected through a correspondingly configured network.

The services' API is implemented in Node.js~\cite{nodejs}, while the PQC algorithms are accessed through the OpenSSL cryptography provider developed in the context of the Open Quantum Safe project~\cite{StebilaMosca2017}. We employed Postgres databases to distinguish between the management of clients', ICA and root CA certificates, so that each ICA is uniquely responsible for the clients' certificates and the root CA is responsible for ICA certificates. Each layer expects the issued certificate to be sent back as an indication of fidelity, otherwise the client will be considered blacklisted, in which case certificate renewal is expected to be trialed in future work. The ICA initiates a new chain

\subsubsection{Software and hardware components}

The main outcome of our implementation concerns an automated certificate issuance system that is divided into three entities -- the root CA, the intermediate CA and the EE. The root CA is located at $\mathit{S}$, the intermediate CA is located at $\mathit{Q_i}$ for some $i$, and the EE is situated at $\mathit{E_{ij}}$ for some $j$. Each authority issues certificates for its adjacent layer from below. We assume that the intermediate CAs with permission to receive a signed certificate are already known to the root CA and kept in a database. Any entity requesting a certificate from the root CA that is not listed in the root CA's record is denied service. The services are written in Node.js~\cite{nodejs} and configured as Docker~\cite{merkel2014docker} images, making the implementation transferable and lightweight.

We distribute our software components among three machines, as shown in Table~\ref{tab:hardware}. Machine 1 was used to run the root CA's and ICAs' services as containerized applications, while Machines 2 and 3 were used to run the ICAS' services and client processes, as described in Section~\ref{sec:edescr}. In the first case of experiments, the framework comprised the two ICAs running on Machines 2 and 3 and, in the second case, the framework comprised one ICA running on Machine 1.

\begin{table*}[htbp]
\centering
\caption{Specifications of the computer machines used for our experiments.}
\begin{tabular}{|c|c|c|c|c|l|}
\hline
\multicolumn{5}{|c|}{\textbf{Hardware Components}} \\
\hline
 \textbf{Name} & \textbf{System Model} & \textbf{Processor} & \textbf{RAM} & \textbf{Operating Software} \\ \hline
Machine 1 & Alienware Aurora R15 AMD & AMD Ryzen 9 7950X 4.5GHz 16 Core          & 64GB                     & Windows 11 Pro     \\ \hline
Machine 2 & Inspiron 5577                             & Intel(R) Core(TM) i7-7700HQ 2.8GHz 4 Core & 24GB                     & Ubuntu 24.04.1 LTS \\ \hline
Machine 3 & HP ZBook Firefly 15 G7 Mobile Workstation & Intel(R) Core(TM) i5-10210U 1.6GHz 4 Core & 16GB                     & Ubuntu 20.04.6 LTS \\ \hline
\end{tabular}
\label{tab:hardware}
\end{table*}

\section{Results and analysis} \label{sec:results}

In this section, we present the results of our experiments conducted as described in Section~\ref{sec:edescr}. We identified four main factors affecting the obtained results, including certificate size, network load, algorithm complexity, but also caching, which signals good functionality of the framework. We consider operations that include digital signatures produced using PQC algorithms at some part of the chain-of-trust and measure the average time taken to complete them. These operations are as follows.
\begin{itemize}
    \item ICA certificate issuance, including:
    \begin{itemize}
        \item ICA download: This action encompasses the transmission of the ICA's CSR, the application of the PQC signature algorithm from the root CA and the reply to the requesting ICA.
         \item ICA upload: This action concludes the ICA's certificate issuance by sending the newly received certificate back to the root CA.
    \end{itemize}
    \item client certificate issuance, including:
    \begin{itemize}
        \item ICA verify: This action initiates the issuance of the client's certificate by confirming with the root CA that the ICA certificate is valid in order to be used. 
         \item client download: This action encompasses the transmission of the client's CSR, the application of the PQC signature algorithm from the ICA and the reply to the requesting client.
         \item client upload: This action concludes the client's certificate issuance by sending the newly received certificated back to the ICA.
    \end{itemize}
    \item root CA application of the signature algorithm for ICA certificate issuance.
    \item root CA verification of the ICA's certificate.
\end{itemize}

\begin{figure}
    \centering
    \includegraphics[width=.85\linewidth]{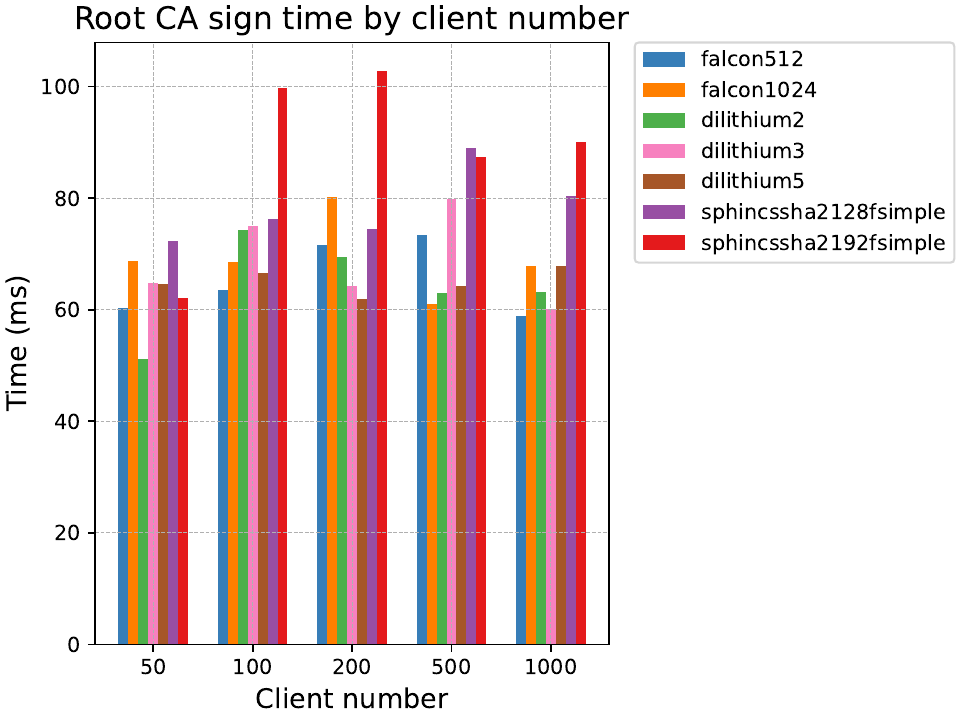}
    \caption{Average time measured for the application of PQC signature algorithms at the root CA level, with respect to the number of EEs (clients). Each bar corresponds to a different PQC algorithm. The certificates signed concern public keys produced with RSA.}
    \label{fig:sign}
\end{figure}

\begin{figure}
    \centering
    \includegraphics[width=.9\linewidth]{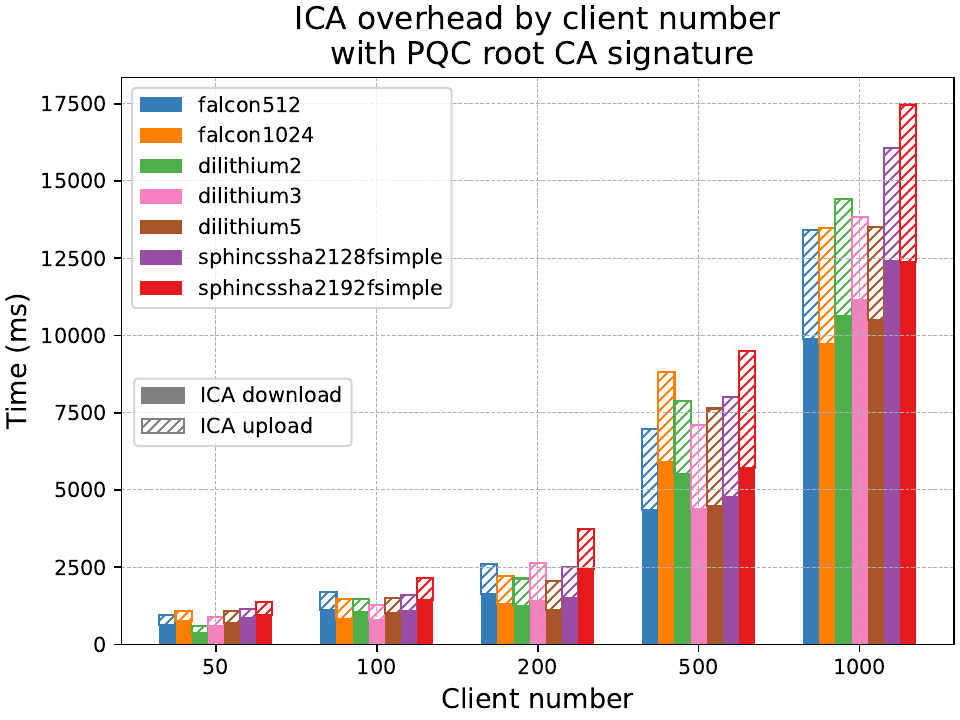}
    \caption{Average time measured for the issuance of certificates to two ICAs, with respect to the number of EEs. Each ICA issued certificates to the same number of EEs. Each bar corresponds to a different PQC signature algorithm employed by the root CA.}
    \label{fig:enroll}
\end{figure}

\begin{figure}
    \centering
    \includegraphics[width=.85\linewidth]{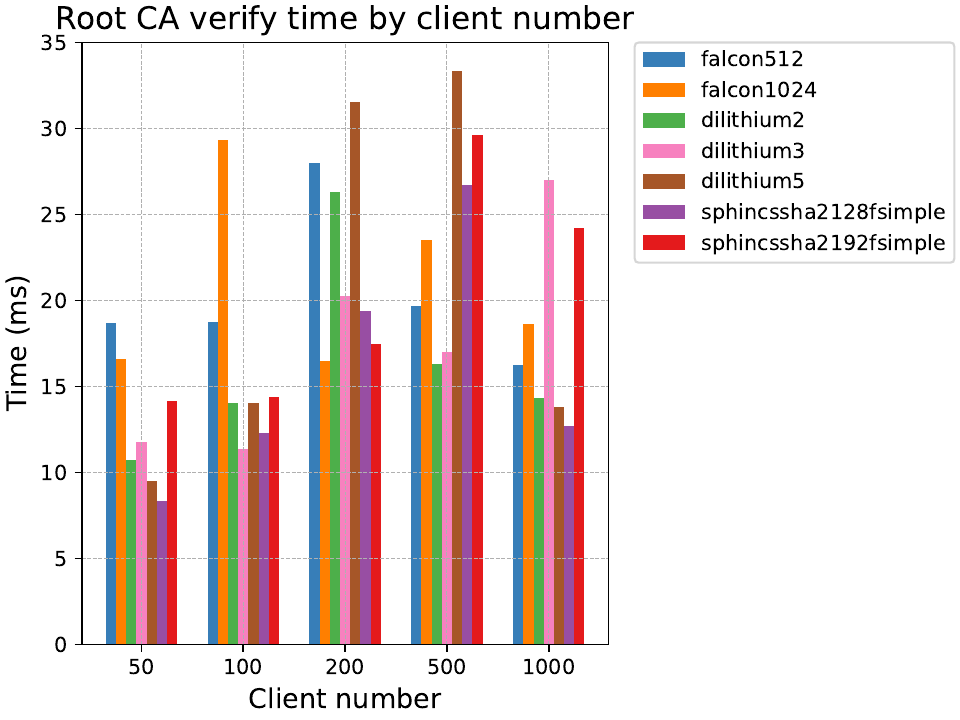}
    \caption{Average time measured for the verification of certificates bearing the root CA PQC signature, with respect to the number of EEs (clients). Each bar corresponds to a different PQC algorithm. The certificates signed concern public keys produced with RSA.}
    \label{fig:verify}
\end{figure}

We observe across all experiments that SPHINCS$^+$ with the larger key size is consistently slow when signing, shown in \textbf{Fig.~\ref{fig:sign}}. This justifies the fact that the employment of the same algorithm places a considerable overhead in the ICAs' certificate issuance time, as seen in \textbf{Fig.\ref{fig:enroll}}. The same does not hold for the verification time with this algorithm, which remains relatively low for up to 200 clients in the first set of experiments and even accelerates as the number of clients increases for the second set of experiments. The results can be seen in \textbf{Fig.~\ref{fig:verify}} This result is owed to the way the algorithm is constructed, as the public key in SPHINCS$^+$ often includes precomputed hash structures and is stateless, meaning no state needs to be tracked between signature verifications. 

\begin{figure}
    \centering
    \includegraphics[width=.9\linewidth]{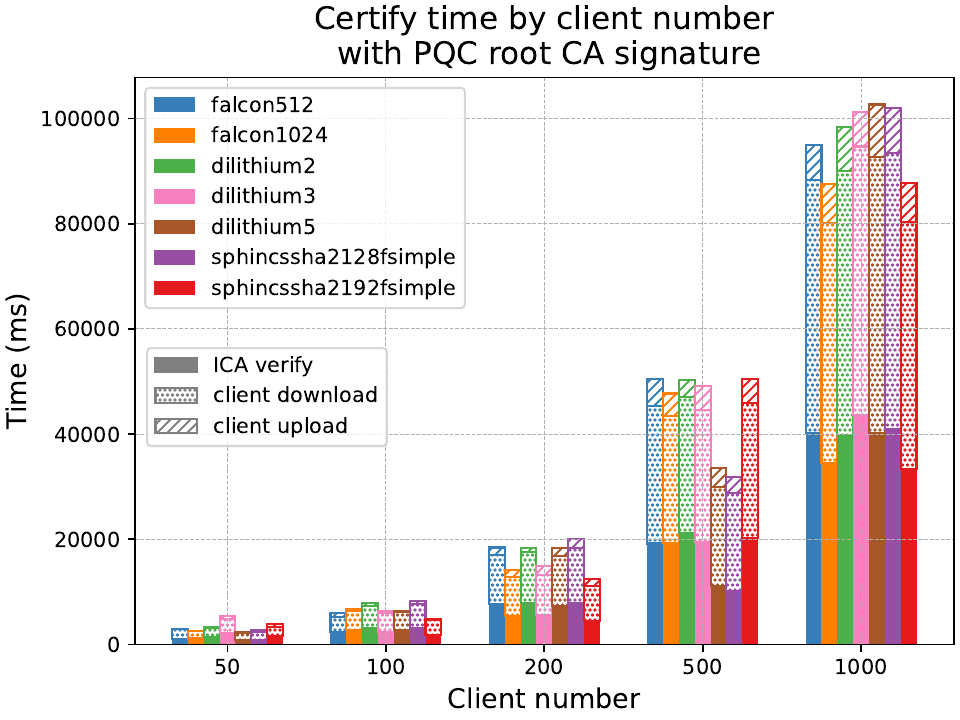}
    \caption{Average time measured for the issuance of certificates to EEs (clients) from two ICAs, with respect to the number of EEs. The EEs were equally distributed to the two ICAs, whose certificates were issued by the root CA. Each bar corresponds to a different PQC signature algorithm employed by the root CA.}
    \label{fig:clientcrt}
\end{figure}

In the case of the experiments with PQC at the root CA level only, we notice that operations including the Falcon variant with the larger public key are generally quicker than employing the ``lighter'' Falcon variant, in terms of measuring the time that a client needs to obtain a signed certificate, as shown in \textbf{Fig.~\ref{fig:clientcrt}}. This is probably due to efficient verification of the ICA certificate, as well as caching and implementation optimizations motivated by the larger key size. We also observe that operations using Falcon512 compete with the times of these using Dilithium2 in terms of average time taken for a client to obtain a signed certificate, while operations using Dilithium3 conclude quicker than when using Dilithium2 for up to 500 clients. Moreover, Dilithium5 and SPHINCS$^+$128 appear unexpectedly efficient, but place considerable overhead when issuing certificates to a number of clients greater than 500.

\begin{figure}
    \centering
    \includegraphics[width=.9\linewidth]{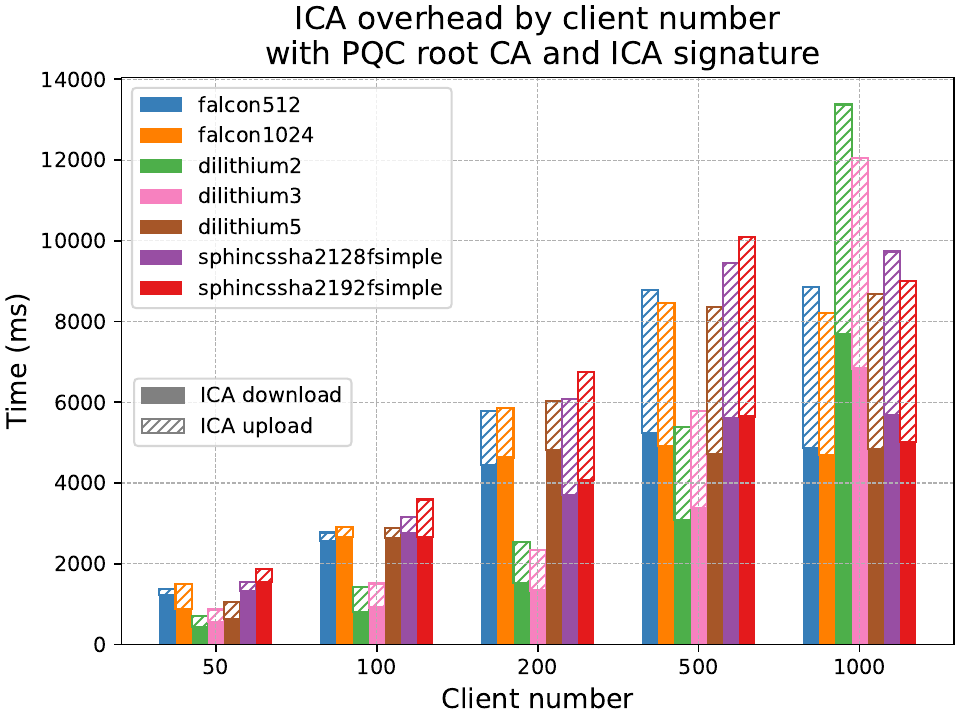}
    \caption{Average time measured for the issuance of certificates to one ICA, with respect to the number of EEs. Each bar corresponds to a different PQC signature algorithm employed by the root CA and the ICA.}
    \label{fig:enroll2}
\end{figure}

\begin{figure}
    \centering
    \includegraphics[width=.9\linewidth]{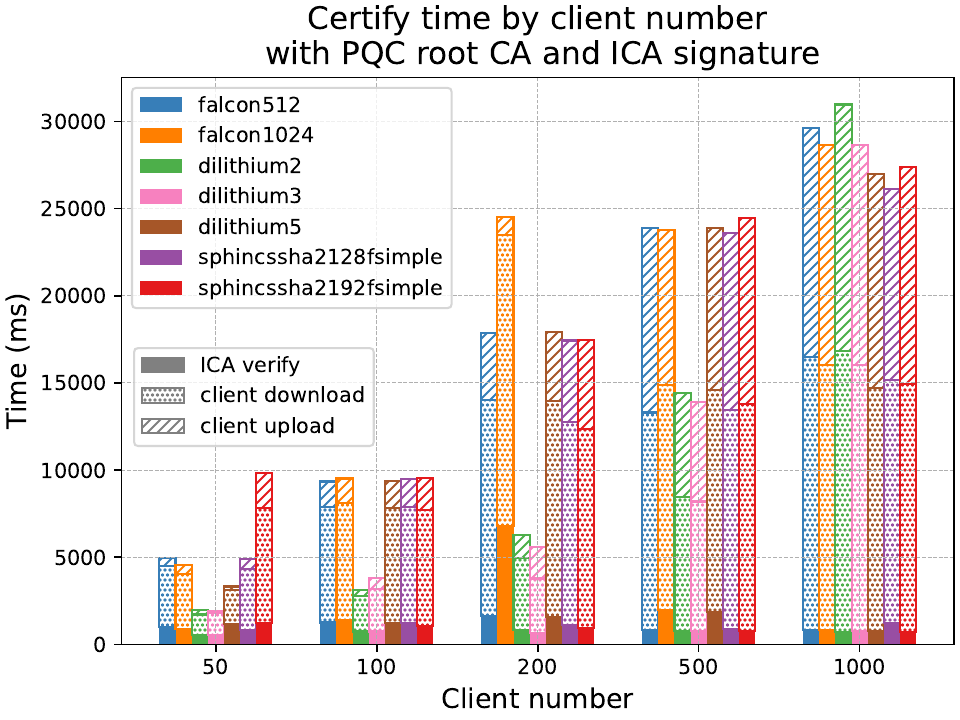}
    \caption{Average time measured for the issuance of certificates to EEs (clients) from one ICA, with respect to the number of EEs. Each bar corresponds to a different PQC signature algorithm employed by the root CA and the ICA.}
    \label{fig:client2}
\end{figure}

In the case of the experiments where PQC algorithms are used both at the root CA and ICA levels, we observe an overall greater overhead in the certificate signing and verification average times, however, the ICA and client certificate issuance operations exhibit an overall acceleration.  We also notice that Dilithium2 and Dilithium3 contribute to a significant increase in overhead when the client numbers range from 500 to 1000 clients for the clients' (\textbf{Fig.~\ref{fig:client2}}) and ICAs' (\textbf{Fig.~\ref{fig:enroll2}}) certificate issuance. In fact, the ICA certificate issuance time decreases with signature algorithms that produce large certificates, contrary to the cases of Falcon, Dilithium2 and Dilithium3, probably due to triggering caching procedures in the framework. 

Overall, we can confirm that SPHINCS$^+$ admits a disadvantage in signing overhead, particularly affecting the average time of issuing ICA certificates from the root CA. However, the results of our application show that this is mitigated by the workflow we design, since the computational burden is placed on verifying certificates at PKI levels usually comprising heavy-duty computers. This renders the SPHINCS$^+$ variants a sensible choice, especially when considering SPHINCS$^+$' properties as an algorithm based on long-studied mathematical foundations. Conversely, Falcon1024 outperforms Falcon512 in client certificate issuance times, likely due to caching optimizations and efficient verification mechanisms, challenging the assumption that smaller key sizes always lead to faster operations. The Dilithium algorithms present mixed results; Dilithium3 generally outperforms Dilithium2 for workloads of up to 500 clients, but as the number of clients grows beyond this threshold, both algorithms encounter significant overhead. Variants like Dilithium5 and SPHINCS$^+$128 face scalability challenges as client numbers increase as well, highlighting the critical impact of certificate size and caching behavior on system performance. These findings underscore the importance of balancing algorithm selection, system design, and workload distribution to optimize certificate authority performance in post-quantum environments.

\section{Conclusion}
In this work, we have presented a scalable framework designed to manage chains-of-trust for a large client base using signature algorithms that include PQC methods. The system accommodates both classical and PQC algorithms, ensuring backward compatibility, adaptability and seamless integration at the critical level of CAs. The framework encompasses services for automated certificate issuance and verification, leveraging PQC signatures for robust protection against quantum-based threats and promoting crypto-agility. The architecture employs a hierarchical three-layer structure comprising a root CA, ICAs and EEs, enabling efficient certificate management. The framework is evaluated in a full-fledged network simulation, demonstrating its performance in signing, verifying, and distributing certificates to multiple clients, with support for up to two ICAs in the hierarchy. The vast usability of the framework's concept motivates its introduction to more diverse network topologies. However, the adaptation to different hardware architectures requires time and effort. We are currently working on including diverse architectures and protocols, while building a versatile, powerful system. Future work includes enhancing automated certificate issuance to meet all ACME protocol requirements, scaling our framework for real-world QKD node domains needing PQC certificates, and testing mixed PQC certificate chains with hybrid signature schemes and key encapsulation mechanisms.
\bibliography{references}

\end{document}